\documentclass[aps,prl,twocolumn,superscriptaddress, showpacs]{revtex4-1}

\usepackage{graphicx}
\usepackage{bm}
\usepackage{amssymb}
\usepackage{pifont}
\usepackage{times}
\usepackage{amsmath}
\usepackage{amsfonts}

\begin{document}


\title{Direct Measurement of the Quantum Wavefunction}


\author{Jeff S. Lundeeni}
\email[]{jeff.lundeen@nrc-cnrc.gc.ca}
\affiliation{Institute for National Measurement Standards, National Research Council, 1200 Montreal Road, Ottawa, Canada, K1A 0R6}

\author{Brandon Sutherland}
\author{Aabid Patel}
\author{Corey Stewart}
\author{Charles Bamber}


\date{\today}

\begin{abstract}
Central to quantum theory, the wavefunction is the complex distribution used
to completely describe a quantum system. Despite its fundamental role, it is
typically introduced as an abstract element of the theory with no explicit
definition \cite{Cohen-Tannoudji2006, Mermin2009}. Rather, physicists come to
a working understanding of the wavefunction through its use to calculate
measurement outcome probabilities via the Born Rule \cite{Landau1989}.
Presently, scientists determine the wavefunction through tomographic methods
\cite{Vogel1989,Smithey1993,Breitenbach1997,White1999,Hofheinz2009}, which
estimate the wavefunction that is most consistent with a diverse collection of
measurements. The indirectness of these methods compounds the problem of
defining the wavefunction. Here we show that the wavefunction can be measured
directly by the sequential measurement of two complementary variables of the
system. The crux of our method is that the first\ measurement is performed in
a gentle way (i.e. weak measurement
\cite{Aharonov1988,Ritchie1991,Resch2004,Smith2004,Pryde2005,Mir2007,Hosten2008,Dixon2009,Lundeen2009,Aharonov2010}%
) so as not to invalidate the second. The result is that the real and
imaginary components of the wavefunction appear directly on our measurement
apparatus. We give an experimental example by directly measuring the
transverse spatial wavefunction of a single photon, a task not previously
realized by any method. We show that the concept is universal, being
applicable both to other degrees of freedom of the photon (e.g. polarization,
frequency, etc.) and to other quantum systems (e.g. electron spin-z quantum
state, SQUIDs, trapped ions, etc.). Consequently, this method gives the
wavefunction a straightforward and general definition in terms of a specific
set of experimental operations \cite{Bridgman1927}. We expect it to expand the
range of quantum systems scientists are able to characterize and initiate new
avenues to understand fundamental quantum theory.
\end{abstract}
\maketitle

The wavefunction $\Psi$, also known as the `quantum state', is considerably
more difficult to measure than the state of a classical particle, which is
determined simply by measuring its position $X$ and momentum $P$. 
According to the Heisenberg Uncertainty Principle, in quantum theory a precise
measurement of $X$ disturbs the particle's wavefunction and forces a
subsequent measurement of $P$ to become random. Thus we learn nothing of the
particle's momentum. Indeed, it is impossible to determine a completely unknown wavefunction of single system \cite{Wootters1982}.

Consider instead performing a measurement of $X$ on an ensemble of particles,
all with the same $\Psi$. The probability of getting result $X=x$ is
$\left\vert \Psi(x)\right\vert ^{2}$. Similarly, the probability of $P=p$
would be $\left\vert \Phi(p)\right\vert ^{2}$, where $\Phi(p)$ is the Fourier
transform of $\Psi(x).$ Even these two probability distributions are not
enough to determine $\Psi(x)$ unambiguously (see the 1d phase retrieval
problem \cite{Trebino2002}). Instead, one must reconstruct $\Psi$ by
performing a large set of distinct measurements (e.g. $Q(\theta)=X\cos
(\theta)+P\sin(\theta)$), and then estimating a $\Psi$ that is most compatible
with the measurement results. This method is known as quantum state tomography
\cite{Vogel1989,Smithey1993,Breitenbach1997,White1999,Hofheinz2009}. In
contrast, we introduce a method to measure $\Psi$ of an ensemble\textit{
directly}. Here, by `direct' we mean that the method is free from complicated sets of measurements and computations; the average raw signal originating from where the wavefunction is being probed is simply proportional to its real and imaginary components at that point. The method rests upon the sequential measurement of two complementary
variables of the system.

At the center of the direct measurement method is a reduction to the
disturbance induced by the first measurement. Consider the measurement of an
arbitrary variable $A$. In general, measurement can be seen as the coupling
between an apparatus and a physical system that results in the translation of
a pointer. The pointer position indicates the result of a measurement. In a
technique known as `weak measurement', one reduces the coupling strength and
this correspondingly reduces the disturbance created by the measurement
\cite{Aharonov1988,Ritchie1991,Resch2004,Smith2004,Pryde2005,Mir2007,Hosten2008,Dixon2009,Lundeen2009,Aharonov2010}%
. This strategy also compromises measurement precision but this can be
regained by averaging. The average of the weak measurement is simply the
expectation value $\left\langle \Psi\right\vert A\left\vert \Psi\right\rangle
$, indicated by an average position shift of the pointer proportional to this amount.

A distinguishing feature of weak measurement is that it does not disturb a
subsequent normal (or `strong') measurement of another observable $C$ in the
limit where the coupling vanishes. For the particular ensemble subset that
gave outcome $C=c,$ one can derive the average of the weak measurement of $A$.
In the limit of zero interaction strength, this is called the Weak Value and
is given \cite{Aharonov1988} by,%
\begin{equation}
\left\langle A\right\rangle _{W}=\frac{\left\langle c\right\vert A\left\vert
\Psi\right\rangle }{\left\langle c|\Psi\right\rangle }.
\end{equation}
Selecting a particular subset of an ensemble based on a subsequent measurement
outcome is known as `post-selection', and is a common tool in quantum
information processing \cite{Knill2001,Duan2001}.

Unlike the standard expectation value $\left\langle A\right\rangle $, the Weak
Value $\left\langle A\right\rangle _{W}$ can be a complex number. This
seemingly strange result can be shown to have a simple physical manifestation:
the pointer's position is shifted by Re$\left\langle A\right\rangle _{W}$ and
receives a momentum kick of Im$\left\langle A\right\rangle _{W}$
\cite{Aharonov1990,Lundeen2005,Jozsa2007}. The complex nature of the Weak Value suggests that it
could be used to indicate both the real and imaginary parts of the wavefunction.

Returning to our example of a single particle, consider the weak measurement
of position ($A=\pi_{x}\equiv\left\vert x\right\rangle \left\langle
x\right\vert $) followed by a strong measurement of momentum giving $P=p$. In
this case the Weak Value is
\begin{align}
\left\langle \pi_{x}\right\rangle _{W}  &  =\frac{\left\langle
p|x\right\rangle \left\langle x|\Psi\right\rangle }{\left\langle
p|\Psi\right\rangle }\label{weak value}\\
&  =\frac{e^{ipx/\hbar}\Psi(x)}{\Phi(p)}.
\end{align}
In the case $p=0$, this simplifies to
\begin{equation}
\left\langle \pi_{x}\right\rangle _{W}=k\cdot\Psi(x),
\end{equation}
where $k=1/\Phi(0)$ is a constant (which can be eliminated later by
normalizing the wavefunction). The average result of the weak measurement of
$\pi_{x}$ is proportional to the wavefunction of the particle at $x$. Scanning
the weak measurement through $x$ gives the complete wavefunction. At each $x$,
the observed position and momentum shifts of the measurement pointer are
proportional to Re$\Psi(x)$ and Im$\Psi(x)$ , respectively. In short, by
reducing the disturbance induced by measuring $X$ and then measuring $P$
normally, we measure the wavefunction of the single particle.

As an experimental example, we performed a direct measurement of the
transverse spatial wavefunction of a photon. Considering a photon travelling
along the $Z$ direction, we directly measure the $X$ wavefunction of the
photon, sometimes called the `spatial mode' (see the Supplementary
Discussion). The Wigner function of the spatial mode of a classical beam has
been measured directly but not for a single photon state
\cite{Mukamel2003,Smith2005}.

\begin{figure}
\includegraphics[ angle=270, width=3.375in]{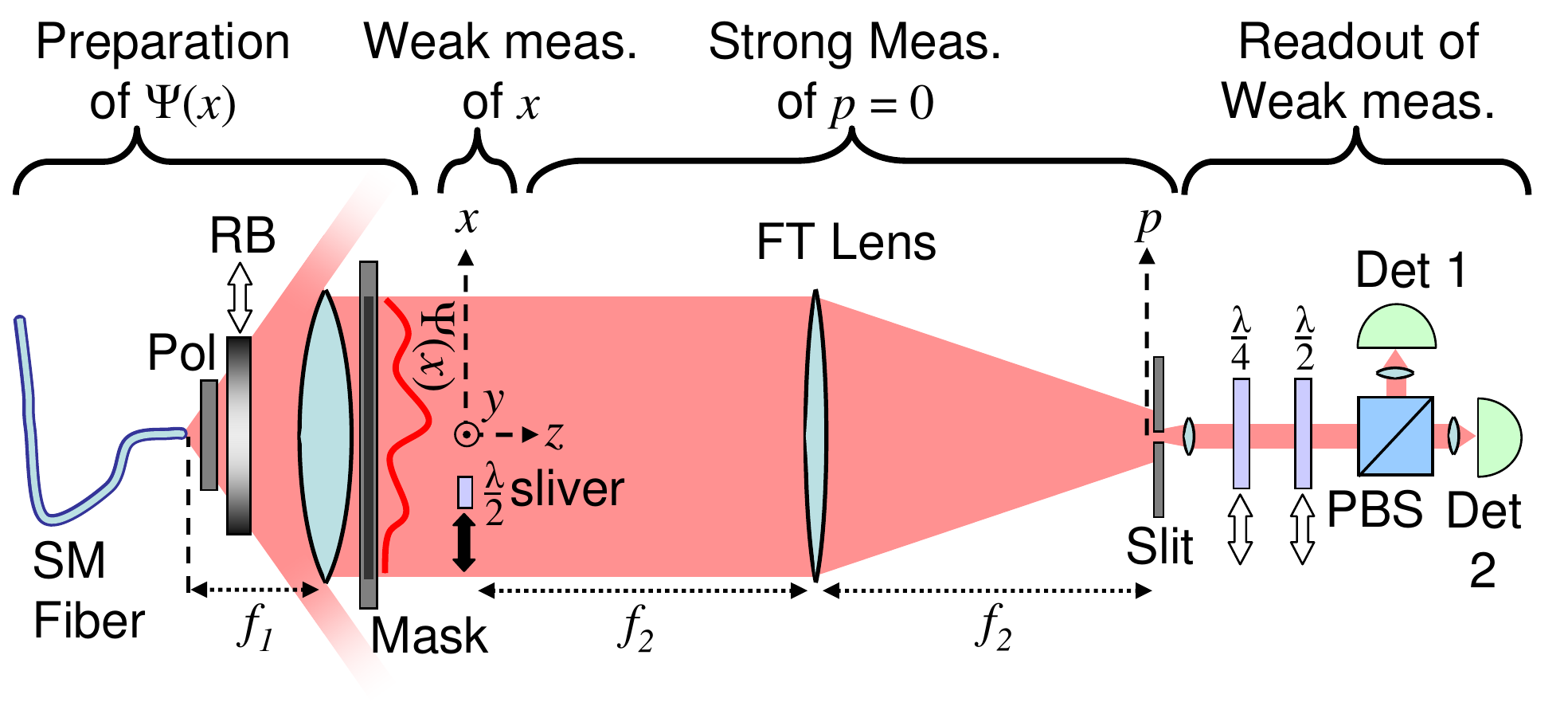}
\caption{\textbf{Direct measurement of the photon transverse wavefunction.} In order to
begin with photons having identical wavefunctions, we transmit them through an
optical fiber that allows only a single mode (SM) to pass. The mode of the SM
fiber (Nufern PM780-HP) is approximately Gaussian with a nominal $1/e^{2}$
diameter of 5.3\ $\pm$ 1.0$\mu$m. The photons emerge from the fiber and pass
through a micro-wire polarizer (Pol) (Edmund Optic NT47-602) to be collimated
by an achromatic lens ($f_{1}=30$cm, diameter=5cm, Thorlabs AC508-300-B), one
focal length $f_{1}$ away from the fiber. The lens was masked off with a
rectangular aperture of dimension $x\times y=43$mm $\times11$mm. Thus our
nominal initial wavefunction was a truncated Gaussian with a $1/e^{2}$
diameter of 56.4mm and a flat phase profile. We modify the magnitude and phase
of the nominal $\Psi(x)$ to create a series of test wavefunctions (see Figs.
3,4). $45$mm past the lens, a rectangular sliver of a half-wave plate
($\frac{\lambda}{2}$ sliver) ($x\times y\times z$ dimensions of $1$%
mm$\times25$mm$\times1$mm) at position $x$ is used to weakly measure $\pi
_{x}=\left\vert x\right\rangle \left\langle x\right\vert $ (see Supplementary
Methods for more detail). The photons then undergo an optical Fourier
transform (FT) induced by an achromatic lens ($f_{2}=1$m, diameter=5cm,
Thorlabs AC508-1000-B), placed one focal length $f_{2}$ from the waveplate
sliver. In the Fourier transform plane, one focal length \emph{$f_{2}$} past
the lens, we postselect those photons with $p=0$ by accepting only those that
pass through a 15 $\mu$m wide slit on axis. We collimate the photons emerging
from the slit with a $f_{3}=3$cm focal length lens. The photons pass through
either a half-wave plate ($\frac{\lambda}{2}$) or quarter-wave plate
($\frac{\lambda}{4}$) and then through a polarizing beamsplitter (PBS). At
each output port, the photons are focused onto a detector (Det 1 and Det 2):
for the single photons, a photon counter (Silicon Avalanche Photodiodes,
PerkinElmer SPCM-AQHR-14); and for the laser, a silicon photodiode (Thorlabs,
DET10A). The imbalance in counts or signal between the two detectors is
proportional to the real ($\frac{\lambda}{2}$) or imaginary part
($\frac{\lambda}{4}$) of the wavefunction. }
\end{figure}

We produce a stream of photons in one of two ways, either by attenuating a
laser beam or by generating single photons through spontaneous parametric
downconversion (SPDC) (see the Supplementary Methods for details). The photons
have a center wavelength of $\lambda=783$nm or $800$nm, respectively. The
experiment (details and schematic in Fig. 1) can be divided into four
sequential steps: preparation of the transverse wavefunction, weak measurement
of the transverse position of the photon, post-selection of those photons with
zero transverse momenta, and readout of the weak measurement.

An ensemble of photons with wavefunction $\Psi(x)$ is emitted from a single
mode (SM) fiber and collimated. We will begin by directly measuring this
wavefunction (described in detail in Fig. 1). We then further test our method
by inducing known magnitude and phase changes to the photons here to prepare a
series of modified wavefunctions.

We weakly measure the transverse position of the photon by coupling it to an
internal degree of freedom of the photon, its polarization. This allows us to
use the linear polarization angle of the photon as the pointer. At a position
$x$ where we wish to measure $\pi_{x}=\left\vert x\right\rangle \left\langle
x\right\vert $ we rotate the linear polarization of the light by $\varphi.$
Consider if $\varphi$ is set to $90^{\circ}$. In this case, one can perfectly
discriminate whether a photon had position $x$ because it is possible to
perfectly discriminate between orthogonal polarizations, $0^{\circ}$ and
$90^{\circ}$. This is a strong measurement. Reducing the strength of the
measurement corresponds to reducing $\varphi$, which makes it impossible to
discriminate with certainty whether any particular photon had $X=x.$ The
benefit of this reduction in precision is a commensurate reduction in the
disturbance to the wavefunction of the single photon.

We then use a Fourier Transform lens and a slit to post-select only those
photons with $p=0$. This constitutes the strong measurement of $P$.

In this subset of the photon ensemble, we find the average value of our weak
measurement of $\pi_{x}$. The average rotation of the pointer, the linear
polarization, is proportional to the real part of the Weak Value. Its
complementary pointer variable, the rotation in the circular polarization
basis, is proportional to the imaginary part of the Weak Value
\cite{Lundeen2005}. Formally, if we treat the initial polarization as a
spin-1/2 spin down vector, then the weak value is given by
\begin{equation}
\left\langle \pi_{x}\right\rangle _{W}=\frac{1}{\sin\varphi}\left(
\left\langle s|\sigma_{x}|s\right\rangle -i\left\langle s|\sigma
_{y}|s\right\rangle \right)  ,
\end{equation}
where $\sigma_{x}$ and $\sigma_{y}$ are the Pauli x and y matrices,
respectively, and $\left\vert s\right\rangle $ is the final polarization state
of the pointer \cite{Lundeen2005}. We measure the $\sigma_{x}$ and $\sigma
_{y}$ expectation values by sending the photons through either through a
half-wave plate or quarter-wave plate, respectively, and then through a
polarizing beamsplitter (PBS). Thus, we read out Re$\Psi(x)$ (half-wave plate)
and Im$\Psi(x)$ (quarter-wave plate) from the signal imbalance between
detectors 1 and 2 at the outputs of the PBS.

\begin{figure}
\includegraphics[width=3.375in]{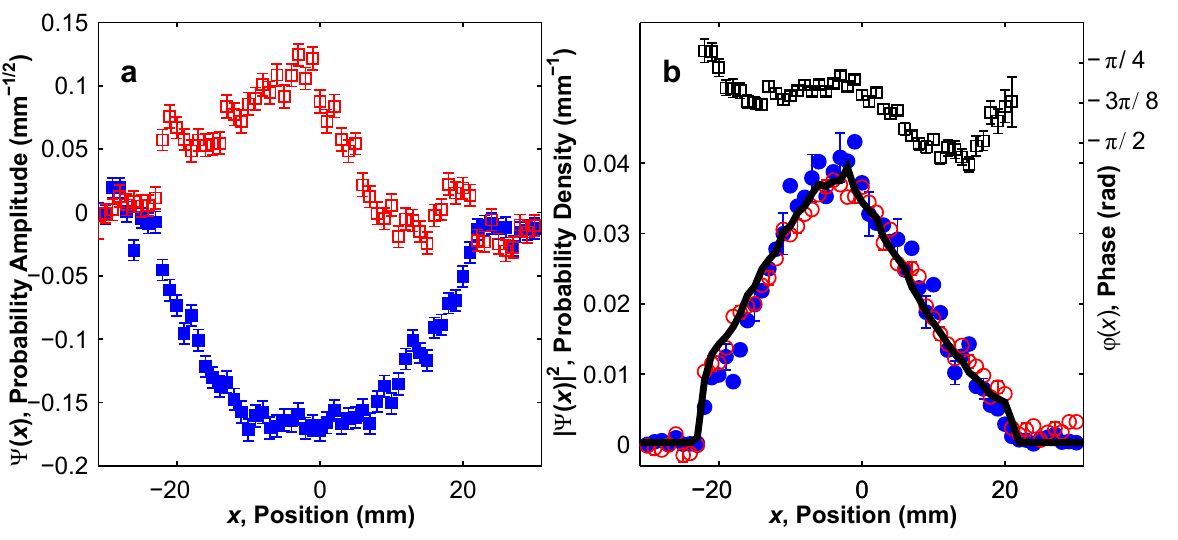}
\caption{\textbf{The measured single-photon wavefunction,} $\Psi(x)$. \textbf{a,}
Re$\Psi(x)$ (solid blue squares) and Im$\Psi(x)$ (open red squares) measured
for the truncated Gaussian wavefunction. \textbf{b,} Using the data in A we
plot the phase $\phi(x)=\arctan\left(  \mbox{Re}\Psi(x)/\mbox{Im}\Psi
(x)\right)  $ (open squares) and the modulus squared $\left\vert
\Psi(x)\right\vert ^{2}$ (open red circles). There is good agreement between
the latter and a strong measurement of the x probability distribution
$\mathrm{Prob}(x)$ (solid line) conducted by scanning a detector along x in
the plane of the sliver. The phase is relatively flat, as expected from the
fiber mode. The slight variation is consistent with the manufacturer
specification of the first lens and the phase curvature measured with a shear
plate. We also removed the slit completely. In this case, there is no
post-selection and the Weak Value $\left\langle \pi_{x}\right\rangle $ becomes
equal to the standard expectation value $\left\langle \Psi\right\vert \pi
_{x}\left\vert \Psi\right\rangle =\left\vert \Psi(x)\right\vert ^{2}.$ We plot
the measured Re$\left\langle \pi_{x}\right\rangle $ (open red circles) after
it is normalized so that$\int\mbox{Re}\Psi(x)dx=1$ and find it is in good
agreement with $\mathrm{Prob}(x)$. We find that Im$\left\langle \pi
_{x}\right\rangle $ is ten times smaller, making $\left\langle \pi
_{x}\right\rangle $ largely real, as expected. Every third error bar (from
statistics) is shown. }
\end{figure}

With $\varphi=20^{\circ}$, we scan our measurement of $\pi_{x}$ in 1 mm steps
and find the Weak Value $\left\langle \pi_{x}\right\rangle _{W}$ at each step.
In this way, we directly measure the photon transverse wavefunction,
$\Psi(x)=\left\vert \Psi(x)\right\vert \exp\left(  i\phi(x)\right)  $. We
normalize the $\sigma_{x}$ and $\sigma_{y}$ measurements by the same factor,
so that $\int\left\vert \Psi(x)\right\vert ^{2}dx=1,$which eliminates the
proportionality constant, $\sin\varphi/\Phi(0).$

To confirm our direct measurement method, we test it on a series of different
wavefunctions. Using our SPDC single photon source, we start by measuring the
initial truncated Gaussian wavefunction (Fig. 2) described in Fig. 1.
Switching to the laser source of photons, we then modify the magnitude, and
then the phase, of the initial wavefunction with an apodized filter and glass
plate, respectively, to create two new test $\Psi$ (Fig. 3). We conduct more
quantitative modification of the wavefunction phase by introducing a series of
phase gradients and then phase curvatures (Fig. 4). For all the test
wavefunctions, we have found good agreement between the expected and measured
wavefunction, including its phase and magnitude (see the Figure Captions for details).

\begin{figure}
\includegraphics[width=3.375in]{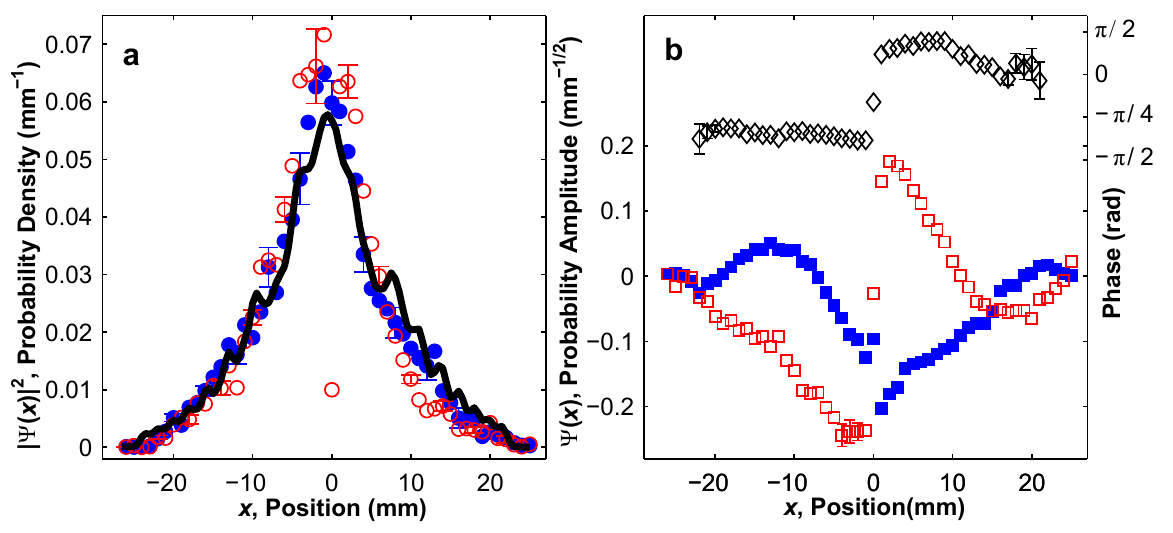}
\caption{\textbf{Measurements of modified wavefunctions.} We further test our ability
to measure $\Psi(x)$ by changing $\mathrm{Prob}(x)$ by placing a reverse
bull's-eye spatially apodized attenuator (RB in Fig. 1) (Edmund Optics, NT64-388) after
the fiber. \textbf{a,} We calculate$\left\vert \Psi(x)\right\vert ^{2}$ from
the data (solid blue circles) along with a detector scan of $\mathrm{Prob}(x)$
(solid line) and find good agreement between the two. \textbf{b,} With the
reverse bull's eye still in place, we modify the phase profile $\phi(x)$ of
the wavefunction by creating a phase discontinuity at $x=0$ imposed with a
glass plate half-way across $\Psi(x)$. At the bottom we show Re$\Psi(x)$
(solid blue squares) and Im$\Psi(x)$ (open red squares), which exhibit a
discontinuity at the plate edge. This discontinuity is even clearer in the
phase difference between the wavefunctions measured with and without the glass
plate, shown at the top (open black diamonds). Despite their discontinuities,
if we use Re$\Psi(x)$ and Im$\Psi(x)$ to calculate$\left\vert \Psi
(x)\right\vert ^{2}$ (\textbf{(a) }open red circles), we find that it is
largely unchanged by the glass plate. This is as expected since the glass has
a transmission near unity. }
\end{figure}

We now describe how the technique of weak measurement can be used to directly
measure the quantum state of an arbitrary quantum system. We have the freedom
to measure the quantum state in any chosen basis $\left\{  \left\vert
a\right\rangle \right\}  $ (associated with observable $A$) of the system. The
method entails weakly measuring a projector in this basis, e.g. $\pi_{a}%
\equiv\left\vert a\right\rangle \left\langle a\right\vert $, and
post-selecting on a particular value $b_{0}$ of the complementary observable
$B$ (See the Supplementary Discussion for a precise definition of
complementarity). In this case, the Weak Value is
\begin{equation}
\left\langle \pi_{a}\right\rangle _{W}=\frac{\left\langle b_{0}|a\right\rangle
\left\langle a|\Psi\right\rangle }{\left\langle b_{0}|\Psi\right\rangle
}=v\cdot\left\langle a|\Psi\right\rangle ,
\end{equation}
where $v$ is a constant, independent of $a$. Thus the Weak Value is
proportional to the amplitude of state $\left\vert a\right\rangle $ in the
quantum state. Stepping $a$ through all the states in the $A$ basis directly
gives the quantum state represented in that basis%
\begin{equation}
\left\vert \Psi\right\rangle =v\cdot\sum\limits_{a}\left\langle \pi
_{a}\right\rangle _{W}\left\vert a\right\rangle ,
\end{equation}
This is the general theoretical result of this paper. It shows that in any
physical system one can directly measure the quantum state of that system by
scanning a weak measurement through a basis and appropriately post-selecting
in the complementary basis.

\begin{figure}
\includegraphics[width=3.375in]{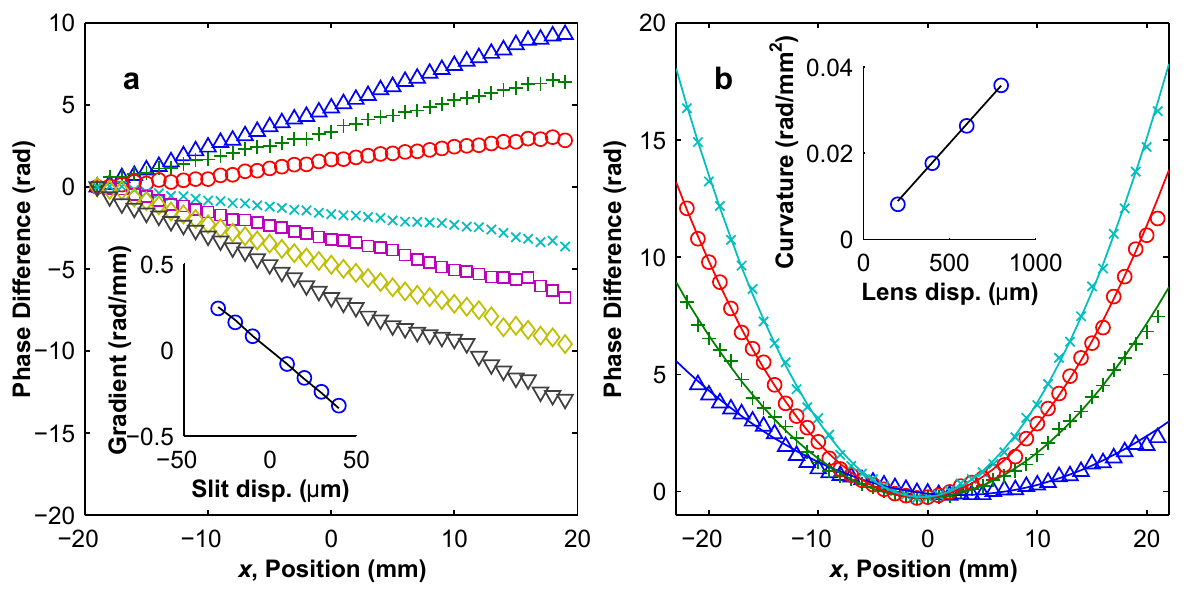}
\caption{\textbf{Phase modification of the wavefunction.} \textbf{a,} We displace the
slit transversely by$\Delta x_{slit}=$-30, -20, -10, 10, 20, 30, and 40 $\mu$m
($\bigtriangleup,+,\bigcirc,\times,\diamond,\square,\bigtriangledown$,
respectively). This effectively redefines the zero momentum axis of the
system. Our photons now travel at an angle to this axis or equivalently the
wavefunction has a linear phase gradient, $\phi(x)=m\cdot x$, where $m=\Delta
x_{slit}2\pi/f_{2}\lambda$. We plot the phase difference between the original
wavefunction and those with a phase gradient. For clarity, the curves have been 
offset to cross at -20mm. This corresponds to shifting the arbitrary global phase of $\Psi(x)$. 
In the inset, is the gradient
$m$ as a function of $\Delta x_{slit}$ (circles) along with theory (line),
which show good agreement. \textbf{b,} We introduce a quadratic phase by
displacing the first lens by $\Delta z=$200, 400, 600, and 800 $\mu$m
($\bigtriangleup,+,\bigcirc,\times$, respectively) along with theoretical fits
(lines). The phase $\phi(x)=r\cdot x^{2}$, where the phase curvature
$r=\pi\Delta z/f_{1}^{2}\lambda$. In the inset, we plot the phase curvature
$r$ from these fits (circles) as a function of lens displacement, $\Delta
x_{slit}$, which shows good agreement with theory (line). Statistical error
bars are smaller than the markers in all the plots. }
\end{figure}

Weak measurement necessarily trades efficiency for accuracy or precision. A comparison of our method to current tomographic reconstruction techniques will require careful consideration of the signal to noise ratio in a given system. In order to increase this ratio in the direct measurement of the photon spatial wavefunction, future experiments will investigate the simultaneous post-selection of many transverse momenta.

In our direct measurement method, the wavefunction manifests itself as shifts
of the pointer of the measurement apparatus. In this sense, the method
provides a simple and unambiguous Operational definition \cite{Bridgman1927}
of the quantum state: it is the average result of a weak measurement of a
variable followed by a strong measurement of the complementary variable. We
anticipate that the simplicity of the method will make feasible the full
characterization of quantum systems (e.g. atomic orbitals, molecular
wavefunctions \cite{Itatani2004}, ultrafast quantum wavepackets
\cite{Dudovich2006}) previously unamenable to it. The method can also be
viewed as a transcription of quantum state of the system to that of the
pointer, a potentially useful protocol for quantum information.

\begin{acknowledgments} This work was supported by the Natural Sciences and
Engineering Research Council and the Business Development Bank of Canada. The concept and theory was by J.S.L. All authors
contributed to the design and building of the experiment and the text of the
manuscript. J.S.L, B.S. and C.B. performed the measurements and data analysis.  Reprints and permissions information is available
at www.nature.com/reprints. The authors declare no competing financial
interests. Readers are weclome to comment on the online version of this
article at www.nature.com/nature. Correspondence and requests for materials
should be addressed to J.S.L. (jeff.lundeen@nrc-cnrc.gc.ca).
\end{acknowledgments}

\bibliographystyle{apsrev4-1}

\section{Supplementary Methods}

\subsection{Single Photon Source}

In this section, we describe how we produce a stream of single photons. A
mode-locked Ti:Sapphire laser produces 100 fs FWHM pulses of light centered
on a wavelength of 800nm (Newport Mai-Tai HP). The pulses are frequency
doubled in a 1mm long BBO crystal cut for Type I collinear phase-matching to
have a wavelength of 400nm. The 800nm red light is reflected away with
dielectric dichroic filters and the 400nm light is focused on a 0.3 mm long
BBO cut for Type II collinear phase-matching. Through spontaneous parametric
downconversion, pairs of collinear photons having a center wavelength $%
\lambda$ of $800$nm are produced in the crystals. One photon is horizontally
polarized (H) and the other is vertically polarized (V). Although they are
produced rarely, since they are always produced in pairs the presence of one
photon can be used to `herald' the presence of its twin. We split the photon
pair into two separate beams with a polarizing beamsplitter (PBS). We couple
the H photon into a multimode fiber which leads to a single photon detector
(Silicon Avalanche Photodiodes, PerkinElmer SPCM-AQHR-13). A click at this
detector then heralds the presence of the V photon in the other beam, thus
producing a single photon state of light. The V photon is the subject of the
measurements we describe in the main article.

\subsection{Laser Source}

In this section we give details of our laser source of photons. We use a
diode laser that produces continuous wave light at $783$nm as a brighter
source of photons with which to test our direct measurement method. We
temperature stabilize the diode to prevent drifts in its wavelength. The
emitted photons are sent through a polarizer and then coupled into a single
mode polarization maintaining fiber, which we couple to the initial single
mode fiber in our experiment.

\subsection{Waveplate Sliver Details}

In addition to the polarization rotation, the waveplate sliver generates two
unwanted transformations in the measured photons, a large $z$ shift and a
phase-shift. Tilting the waveplate sliver about the $x$ axis allows us to
null the induced phase-shift. We precompensate for the $z$ shift before the
fiber in the following way. Because the photon is fairly broadband
($>$10nm), its coherence length is short ($<$30 um). The waveplate sliver ($%
h\approx1$mm thick) described in the main paper displaces the photon by $%
z_{s}=h\cdot(n-1)$, where $n$ is the optical index. This displacement is
longer than the photon's coherence length. If displaced beyond one coherence
length, the photon will not exhibit the required interference between the
amplitude for a photon to go through the sliver and the amplitude to go
around it. Therefore we precompensate for this by first creating two amplitudes
for the single photon separated by exactly the displacement created by
waveplate sliver. We achieve this exactness by using another piece of the
waveplate the from which the sliver was cut. By placing this extra piece of waveplate
so that it covers half of the vertically polarized beam emerging from the PBS and then coupling
that beam into a single-mode polarization maintaining optical fiber (Nufern
PM780-HP), we create the two amplitudes, $T(z=0)$ and $T(z_{s})$. We measure the wavefunction of the photons emerging from the other end of this fiber.

Propagating these two amplitudes through the waveplate sliver, we have $T(z_{s})$ and $%
T(2z_{s}),$ whereas the amplitudes to go around the sliver will remain $T(z=0)
$ and $T(z_{s})$. The $T(z_{s})$ amplitudes from each process will
interfere, whereas the $T(z=0)$ and $T(2z_{s})$, being orthogonal, will not.
The latter amplitudes simply add a constant background to our measurement of 
$\sigma_{x}$ and $\sigma_{y}$ halving the magnitude of $\left\langle
s|\sigma_{x}|s\right\rangle $ and $\left\langle s|\sigma_{y}|s\right\rangle .
$ Nonetheless the proportionality to $\Psi(x)$ remains and this decrease in
signal is inconsequential after normalizing $\Psi(x).$

\section{Supplementary Discussion}

\subsection{The Transverse Photon Wavefunction}

The total quantum state of the photon is a function of the photon's
polarization and the three spatial degrees of freedom, $X$, $Y$, and $Z$. If
there are no correlations between these four degrees of freedom, one can write the
total wavefunction of the photon as a product of wavefunctions, one for each degree of freedom. Under
these conditions, one can measure the wavefunction of each degree of freedom
separately. Considering a photon travelling along the $Z$ direction, we
directly measure the $X$ wavefunction of the photon.

There are techniques for measuring the spatial mode of a
classical beam of light. These include spatial-shear interferometry,
Shack-Hartman sensors \cite{Platt2001}, spiral interferometry
\cite{Juanola-Parramon2008}, the Gerchberg-Saxton algorithm
\cite{Gerchberg1972}, and conoscopic holography \cite{Buse2000}. Some of these
might be applicable to single photons. However, none of these techniques are
direct, either measuring only phase gradients $\partial\phi/\partial x$ or
$\partial^{2}\phi/\partial x^{2}$ and stitching these together to estimate the
actual phase $\phi$, or requiring computer algorithms to interpret the
results.

\subsection{Complementarity}

Central to this method is the concept of complementarity, which is formalized
by the theory of Mutually Unbiased (MU) Bases \cite{Durt2010}. Two bases $%
\left\{ \left\vert a\right\rangle \right\} $ and $\left\{ \left\vert
b\right\rangle \right\} $ are MU if all their constituent states have the
same overlap, i.e. for any values of $a$ and $b,$$\left\vert \left\langle
a|b\right\rangle \right\vert ^{2}=1/N$, where $N$ is the dimension of the
Hilbert space. A strong measurement of $A$ will leave the system in a flat
distribution of the $\left\{ \left\vert b\right\rangle \right\} $ basis. In
any physical system there are always at least two bases that are mutually
unbiased. By unitary transformation, once one of the bases in the MU pair is
selected, the other is then fixed. Thus we are free to choose
our basis for the direct measurement of the quantum state (i.e. we are
guaranteed there will be at least one variable that we can
post-select on). In every basis that is unbiased with respect to our direct measurement basis, there is a
state $\left\vert b=b_{0}\right\rangle $ for which $\left\langle
b_{0}|a\right\rangle $ is a constant, i.e. the phase of the overlap is
independent of $a$ \cite{Durt2010}. Post-selecting on this state will result
in a weak value that that is proportional to the quantum state. The
proportionality constant, $v=\left\langle b_{0}|\Psi\right\rangle
/\left\langle b_{0}|a\right\rangle $, is a constant, independent of $a$.

The choice of state to post-select on is in fact somewhat arbitrary. Any
state $\left\vert b_{k}\right\rangle $ in $\left\{ \left\vert b\right\rangle
\right\} $can be used. Now, the stipulation that$\left\langle
b_{k}|a\right\rangle $is a constant sets the laboratory reference frame relative to which
$\Psi$ is measured. In the measurement of the photon transverse
wavefunction, momentum is defined relative to a lab coordinate system set by
an axis joining the center of the Fourier Transform lens to the slit. One is
thus free to set $p=0$ to be whatever direction one chooses; the
wavefunction will be measured relative to this coordinate frame.

The theory of MU bases is less developed for continuous variables such as
those in our example, $X$ and $P$ of a single particle. In a Hilbert space
defined by an unbounded continuous variables such as these, for any chosen
variable $Q(\theta )=X\cos (\theta )+P\sin (\theta )$ then $Q(\theta ^{\prime
})$ will be respectively unbiased for any $\theta \neq \theta ^{\prime }$;
there exist an infinite number of MU pairs of bases \cite%
{Weigert2008}. Theoretically, this creates a great deal of flexibility in
the method to directly measure the wavefunction. In practice though, basis states
of continuous variables are not physical, in that the range of $Q$ in
projector $\pi _{q}$ is zero, $\Delta q=0.$ For a finite range measurement, $%
\Delta q>0$, the associated states are no longer mutually unbiased. In this
case, the most unbiased pair of bases will be $Q(\theta )$ and $Q(\theta
+\pi /2),$ e.g. $X$ and $P$.


\end{document}